\documentclass{nature}
\usepackage{booktabs,caption,fixltx2e,rotating,epsfig}
\usepackage[flushleft]{threeparttable}
\usepackage{graphicx}

\newcounter{firstbib}

\title{Puzzling accretion onto a black hole in the ultraluminous X-ray source M101 ULX-1}

\author{Jifeng Liu$^{1}$,
Joel N. Bregman$^{2}$,
Yu Bai$^{1}$,
Stephen Justham$^{1}$,
Paul Crowther$^{3}$,
}

\def\lesssim{\mathrel{\hbox{\rlap{\hbox{\lower4pt\hbox{$\sim$}}}\hbox{$<$}}}}
\def\gtrsim{\mathrel{\hbox{\rlap{\hbox{\lower4pt\hbox{$\sim$}}}\hbox{$>$}}}}
\def\ggg{\mathrel{\hbox{\rlap{\hbox{\lower4pt\hbox{$\sim$}}}\hbox{$>$}}}}
\def\persec{s$^{-1}$}

\begin{document}

\maketitle

\begin{affiliations}
\item Key Laboratory of Optical Astronomy, National Astronomical Observatories, Chinese Academy of Sciences, 20A Datun Rd, Chaoyang District, Beijing, China 100012
\item Department of Astronomy, University of Michigan, 500 Church St., Ann Arbor, MI 40185, US
\item Department of Physics \& Astronomy, University of Sheffield, Hounsfield Rd, Sheffield S3 7RH, UK
\end{affiliations}

\begin{abstract}

There are two proposed explanations for ultraluminous X-ray
sources\cite{Fabbiano05,Gladstone13} (ULXs) with luminosities in excess of $10^{39}$ erg
s$^{-1}$.  They could be intermediate-mass black holes (more than
100-1,000, solar masses, $M_\odot$) radiating at sub-maximal (sub-Eddington) rates, as in
Galactic black-hole X-ray binaries but with larger, cooler accretion
disks\cite{Miller04,Kong04,Liu08}.  Alternatively, they could be stellar-mass
black holes radiating at Eddington or super-Eddington
rates\cite{Gladstone09,Gladstone13}.  On its discovery, M101
ULX-1\cite{Kong04,Mukai05} had a luminosity of $3\times10^{39}$ erg
s$^{-1}$ and  a supersoft thermal disk spectrum with an exceptionally low temperature ¡ª- 
uncomplicated by photons energized by a corona of hot electrons -- more consistent with the
expected appearance of an accreting intermediate-mass black
hole\cite{Kong04,Miller04}.
Here we report optical spectroscopic monitoring of M101 ULX-1. We confirm the
previous suggestion\cite{Liu09} that the system contains a Wolf-Rayet star, and
and reveal that the orbital period is 8.2 days. The black hole has a minimum mass of
$5M_\odot$, and more probably a mass of $20-30M_\odot$, but we argue that it
is very unlikely to be an intermediate-mass black hole.
Therefore its exceptionally soft spectra at high Eddinton ratios violate the
expectations for accretion onto stellar-mass black holes\cite{McClintock06,Esin97,rem06}.
Accretion must occur from captured stellar wind, which has hitherto been
thought to be so inefficient that it could not power an ultraluminous
source\cite{Rappaport05,Linden10}.

\end{abstract}

While it is desirable to obtain the primary mass for ultraluminous X-ray
sources (ULXs) through measuring the motion of its companion, this is only
possible in the X-ray low state because the X-ray irradiated accretion disk
will dominate the optical light in the high state\cite{Roberts11,Cseh11}.
A spectroscopic monitoring campaign for M101 ULX-1 was carried out
from February to May 2010 during its expected X-ray low states.
%
The optical spectrum (Figure~1) is characterized by broad helium emission
lines,  including the He II 4686\AA\ line.
Given the absence of broad hydrogen emission lines, which are detected in some
ULXs from their X-ray irradiated accretion disk at very high
luminosities\cite{Roberts11,Cseh11}, the donor cannot be hydrogen rich, and
thus must be a Wolf-Rayet (WR) star or a helium white dwarf. The latter can be
excluded because a white dwarf is roughly a million times dimmer than the
observed optical counterpart even during the low states.
Indeed, the optical spectrum is unique to WR stars, and the intensities of the
helium emission lines can be reproduced well by an atmospheric
model\cite{Hillier98} of a WR star, the mass of which is estimated to be
$19M_\odot$ based on the empirical mass-luminosity
relation\cite{Schaerer92,Crowther07}.
Given the relatively low luminosities in the X-ray low state, the helium
emission lines are expected to originate mainly from the WR secondary with
little contribution from the accretion disk.
Such emission lines
have been used to measure the black hole (BH) mass in both IC10 X-1
($21-35M_\odot$)\cite{Prestwich07,Silverman08} and NGC300 X-1
($12-24M_\odot$)\cite{Carpano07,Crowther10}, systems which exhibit
luminosities an order of magnitude lower than the peak luminosity of M101
ULX-1.

Since the centroid of the He II 4686\AA\ emission line varied by $\pm60$ km/s
over three months of our monitoring campaign, we have been able to obtain the
orbital period of $P = 8.2 \pm 0.1$ days and the mass function
$f(M_*,M_\bullet,i)  = 0.18 \pm 0.03 M_\odot$ for M101 ULX-1 (Figure~2).
Because we already know the mass of the donor star we are able to infer the
mass of the accretor to be $M_\bullet \ge 4.6 \pm 0.3 M_\odot$ (for inclination
angle $i \le 90^\circ$), where the error is computed from the uncertainties in
the secondary mass and in the mass function.
Even for the minimum mass, obtained when the system is aligned perfectly
edge-on to the line of sight (for which $i=90^\circ$), such a compact primary
can only be a BH.
Higher BH masses are easily obtained for lower inclination angles.  For
example, a stellar mass BH of $20M_\odot$ corresponds to $i=19^\circ$, and an
intermediate mass black hole (IMBH) of $1000M_\odot$ ($300M_\odot$) corresponds to $i=3^\circ$ ($i=5^\circ$).
The probability of discovering a pole-on binary with $i < 3^\circ$ ($i=5^\circ$)
by mere chance is lower than 0.1\% (0.3\%). This makes it very unlikely that
this system contains an IMBH of $1000M_\odot$ ($300M_\odot$).
If the peak luminosity of M101 ULX-1 corresponds to less than 30\% of the
Eddington level -- which is commonly-assumed to be required to produce the
thermally-dominated spectral state\cite{McClintock06,Steiner09} -- then the BH
mass would exceed $50-80M_\odot$.
The true BH mass seems likely to be $\sim20-30M_\odot$ (see the Supplementary
Information for details).


The confirmation of a WR star in the system, independent of the dynamical mass
measurement, also suggests that M101 ULX-1 is unlikely to be an IMBH.
IMBHs cannot form directly through the collapse of massive stars, but it is
suggested that they can form through mergers in dense stellar
environments\cite{Miller02,Portegies04}.
However, any IMBH formed would not be seen as a ULX unless they capture a
companion as a reservoir from which to accrete matter.
Such a capture is a rare event even in dense stellar environments such as
globular clusters or galactic bulges, to which M101 ULX-1 apparently does not
belong, and captures that can provide high-enough accretion rates to power a
ULX are even more unusual\cite{Blecha06,Madhusudhan06}.
Given the rarity of WR stars, with roughly 2000 WR stars out of 200 billions of
stars in a typical spiral galaxy like the Milky Way\cite{Crowther07}, it is
extremely unlikely that M101 ULX-1 is such a revived IMBH.
Alternatively a huge population of IMBHs would somehow remain undetected, both
with and without companions.
%


M101 ULX-1 is thus a stellar black hole, although it is a member of the class
of supersoft ULXs which have been considered to be outstanding IMBH
candidates\cite{Kong04,Liu08}.
Its combination of high luminosities and low disk temperatures (Figure~3)
strains our current understanding of accretion by stellar-mass
BHs\cite{Esin97,McClintock06,rem06}.
Studies of Galactic black hole X-ray binaries suggest that radiation at less
than roughly 30\% of the Eddington luminosity is dominated by the thermal
emission from a hot disk ($\sim1$ keV). A hard power-law component due to
Comptonization by the disk corona becomes more and more significant when the
luminosity increases to near-Eddington levels.
When the luminosity increases further to Eddington or super-Eddington levels,
the Comptonized component begins to dominate over the disk component, as
observed for ULXs in the ultraluminous state\cite{Gladstone09,Gladstone13}.
For example, the ultraluminous microquasar in M31 with a stellar-mass black
hole ($\sim10M_\odot$) and a luminosity of $10^{39}$ erg s$^{-1}$ exhibited
hard X-ray spectra\cite{Middleton13}.
If it were the same phenomenon, a hard X-ray spectra would be expected for a
stellar-mass BH in M101 ULX-1, whether it is radiating at sub-, near- or
super-Eddington luminosities.
The observed supersoft X-ray spectra lack hard photons above 1.5 keV, and can
be described purely by cool accretion disks, uncomplicated by Comptonization,
with exceptionally low temperatures of 90-180 eV\cite{Kong04,Mukai05}.
Including extra photoelectric absorption by the surrounding WR wind into
spectral analysis would further lower the underlying disk temperatures and
increase the luminosities\cite{Kong04}, which would drive M101 ULX-1 to deviate
even farther from the expected hard spectra.
%
This unambiguously demonstrates that stellar mass BHs can have very cool
accretion disks uncomplicated by the Comptonized component, contrary to
standard expectations\cite{Miller04,McClintock06,rem06} .

%
%

M101 ULX-1 is the third known WR/BH binary but is distinctly different from NGC
300 X-1 and IC 10 X-1.
While M101 ULX-1 is a recurrent transient with
supersoft spectra and low disk temperatures, both IC 10 X-1 and NGC 300 X-1
show constant X-ray output
(despite apparent variations due to orbital modulation), hard spectra with a
minor disk component, and disk temperatures above 1
keV\cite{Prestwich07,Carpano07,Barnard08} (Figure~3).
Hence the compact object in M101 ULX-1 was considered to be an excellent IMBH
candidate, while IC 10 X-1 and NGC 300 X-1 were expected to host stellar mass
black holes (as was later confirmed).
The 8.2-day orbital period shows that M101 ULX-1 is a wide binary, with
components which would be separated by $50R_\odot$ for $M_\bullet = 5M_\odot$
($75R_\odot$ for $M_\bullet = 60M_\odot$).
The Roche lobe radius for the secondary is always greater than $22R_\odot$,
twice as large as the WR star itself. Mass transfer by Roche lobe overflow is
thus impossible, and the black hole must be accreting matter by capturing the
thick stellar wind.
Given the geometry of the system, the disk is very large, and thus there will
be a helium partial ionization zone.  Such a disk is prone to instability,
causing the observed X-ray transient behaviors for M101 ULX-1.
In contrast, IC 10 X-1 and NGC 300 X-1 have shorter orbital periods (34.9 hr
and 32.3 hr respectively) and smaller separations ($\sim20R_\odot$). Since
those WR stars fill their Roche lobes, the BHs accrete via Roche-lobe overflow.
These systems also have much smaller and hotter accretion disks without helium
partial ionization zones,  which explains why IC 10 X-1 and NGC 300 X-1 do not
display disk-instability outbursts (see also the Supplementary Information).


Mass transfer through wind-accretion usually has a very low efficiency, as in
the case of many low-luminosity high mass X-ray binaries, and is typically not
considered for populations that require high accretion rates.
However, M101 ULX-1 demonstrates that this expectation is not always correct.
In particular, transient outbursts of such wind-accreting system have generally
not been included in theoretical ULX populations \cite{Rappaport05,Linden10},
but M101 ULX-1 does attain ULX luminosities.
%
%
Theorists have recently suggested that wind accretion may potentially also be
significant for some progenitors of type Ia supernovae\cite{Mohamed11}. M101
ULX-1 empirically supports this reassessment of the potential importance of
wind accretion.


%
%
%
%

\begin{addendum}

 \item[Acknowledgements] We thank Drs. Jeffery McClintock, Rosanne Di Stefano,
Qingzhong Liu, Xiangdong Li, Feng Yuan, and ShuangNan Zhang for helpful
discussions.  J.F. Liu acknowledges support for this work provided by NASA
through the Chandra Fellowship Program (grant PF6-70043) and support by
National Science Foundation of China through grant NSFC-11273028.  The paper is
based on observations obtained at the Gemini Observatory, which is operated by
the Association of Universities for Research in Astronomy, Inc., under a
cooperative agreement with the NSF on behalf of the Gemini partnership: the
National Science Foundation (United States), the National Research Council
(Canada), CONICYT (Chile), the Australian Research Council (Australia),
Minist\'{e}rio da Ci\^{e}ncia, Tecnologia e Inova\c{c}\~{a}o (Brazil) and
Ministerio de Ciencia, Tecnolog\'{i}a e Innovaci\'{o}n Productiva (Argentina).

 \item[Author contributions] J.F. Liu and J.N. Bregman proposed the
observations, J.F. Liu and Y. Bai reduced the data and carried out the
analysis, J.F.Liu, J.N. Bregman and S. Justham discussed the results and wrote
the paper, P. Crowther helped to confirm the properties of the Wolf-Rayet star.
All commented on the manuscript and contributed to revise the manuscript.


 \item[Author information] Reprints and permissions information is available at
npg.nature.com/reprints. The authors declare that they have no competing
financial interests. Readers are welcome to comment on the online version of
the paper. Correspondence and requests for materials should be addressed to
J.F. Liu  (email: jfliu@nao.cas.cn).

\end{addendum}

\newpage

{\noindent {\bf Figure~1: The secondary of M101 ULX-1 is confirmed to be a
Wolf-Rayet star based on the optical spectrum}, combined from 10 Gemini/GMOS
observations with a total exposure time of 16 hours. The spectrum shows narrow
nebular lines with FWHM of $\sim$4\AA\ at the instrumental spectral resolution,
including hydrogen Balmer lines and forbidden lines such as [OIII] 4960/5006
(the latter is mostly in the CCD gap and only partly shown), [NII] 6548/6583,
and [SII] 6716/6731, all at a constant radial velocity over observations
consistent with that of M101.  Also present are broad emission lines  with FWHM
up to 20\AA, including strong HeII 4686, HeI 5876, HeI 6679, weaker HeI 4471,
HeI 4922, and HeII 5411, and NIII 4640 lines.
%
The observed HeI 5876/He II 5411 equivalent width ratio suggests a WR star of
WN8 sub-type, consistent with the absence of carbon emission lines for WC stars
(such as CIII 5696 and CIV 5812).
The intensities of the helium emission lines can be best reproduced by an
atmospheric model\cite{Hillier98} of a Wolf-Rayet star with $R_*=10.7R_\odot,
M_* = 17.5 M_\odot, L_* = 5.4\times10^5 L_\odot, T_* = 48 {\rm kK}, \dot{M}_* =
2 \pm 0.5\times10^{-5} M_\odot/yr, v_\infty = 1300\pm100$ km \persec\ (with
68.3\% uncertainties for the two continuously variable parameters), consistent
with those for a WN8 star.
The mass-luminosity relation\cite{Schaerer92,Crowther07} for WR stars gives a
more reliable mass estimate of $19M_\odot$, which we use in the main text, with
an estimated formal error of $1M_\odot$. }



\newpage

{\noindent {\bf Figure~2: An orbital period of $\sim8.2$ days is revealed from
the radial velocity measurements over three months for M101 ULX-1.}
The upper panel (a) shows the radial velocities of the HeII 4686\AA\ emission line
(with 68.3\% uncertainties computed mainly from the dispersion of the
wavelength calibration) from nine observations over three months.  The lower
right panel (b) shows the $\chi^2$ computed for a sine fit (under the assumption of
a circular orbit) to the radial velocity curve as a function of trial periods.
The trial periods range from a minimum of 3 days, when the WR secondary fills
its Roche lobe, to a maximum of 10 days as suggested by the last five
measurements. The best fit is achieved at minimal $\chi^2 \sim1.6$ for $P=8.2$
days and $K=61$ km/s, for which the folded radial velocity curve is shown in
the lower left panel (c). The 68.3\% uncertainties for the best fit are estimated
to be $\Delta P = 0.1$ days and $\Delta K = 5$ km/s using $\chi^2 -
\chi^2_{best} = 1$. All other trial periods (such as those at $P \sim 6.4$
days) are worse by $\Delta\chi^2>4$.
The successful fit with a sine curve suggests that the orbital eccentricity is
small.
This leads to a mass function $f(M_*,M_\bullet,i) = {P K^3 \over 2\pi G} = 0.18
\pm 0.03 M_\odot$, where the error accounts for the 68.3\% uncertainties in $P$
and $K$.  }

\newpage

{\noindent {\bf Figure~3:
The prototype ultraluminous supersoft X-ray source M101 ULX-1 exhibited
distinct spectral characteristics. } M101 ULX-1 is compared to Galactic black
hole X-ray binaries (GBHXRBs), WR/BH binaries IC 10 X-1 and NGC 300 X-1, and
other ULXs on the disk X-ray luminosity ($L_X$) - temperature ($T_d$) plane,
all plotted with the 68.3\% uncertainties from the X-ray spectral fitting.
Except for M101 ULX-1, which can be fitted with a disk blackbody model with
temperatures of 90-180 eV\cite{Kong04,Mukai05}, all other X-ray sources are
complicated by the presence of a hard power-law component due to Comptonization
by a corona, and can be best fitted with a disk blackbody plus power-law
composite model\cite{Miller04,Barnard08}.
While GBHXRBs\cite{Miller04} and the other two WR/BH binaries\cite{Barnard08}
with stellar black holes cluster in the same region, M101 ULX-1 lies within a
distinct region that has been expected to contain IMBH candidates, the same
region as for some ULXs\cite{Miller04}.
The dotted lines describe the expected disk luminosity ($L_d$) for different
disk temperatures  for a fixed disk inner radius based on the relation $L_d
\propto R_{in}^2 T_d^4$. The two lines are offset by 4 orders of magnitudes in
luminosity, implying a factor of 100 difference in the disk inner radii, and a
factor of 100 difference in the black hole masses if the disk radius is tied to
the innermost stable orbit of the black hole.
Fitting ULX spectra with alternative Comptonization models can yield high disk
temperatures consistent with those of stellar mass black
holes\cite{Gladstone09}.
However, the location of M101 ULX-1 on the $L_X-T_d$ plane does not change
because its spectra are not complicated by Comptonization at all.
}

\newpage

{\noindent {\bf Extended Data Figure 1. M101 ULX-1 as observed in the optical.}
(a) M101 ULX-1 is located on a spiral arm of the face-on grand-design spiral
galaxy M101, as indicated by the arrow. The color image of M101 is composed of
GALEX NUV, SDSS g, and 2MASS J images.  (b) ULX-1 is identified as a blue
object with V=23.5 mag at the center of the $1^{\prime\prime}$ circle on the
HST image. The color image is composed of ACS/WFC F435W, F555W and F814W
images. }

{\noindent {\bf Extended Data Figure 2. Physical properties of the WR secondary
from spectral line modeling.} Distributions of computed $\Delta^2$ as a
function of (a) stellar masses, (b) stellar mass loss rate, (c) stellar radii,
and (d) terminal velocity. Here $\Delta^2 = \sum_i ({\rm EW} - {\rm EW}_i)^2$
computes the difference between observed and synthetic equivalent widths for
six broad helium lines present in the Gemini/GMOS spectrum. We have computed
synthetic spectra for a group of 5000 real stars from the evolution tracks (as
shown by the thick stripes in the mass plot and the radius plot) and for
another group of "fake" stars with continuous distributions in mass, radius and
luminosity.  The best model is labeled by a filled pentagon in all panels.  }

{\noindent {\bf Extended Data Figure 3. Properties for the Wolf-Rayet/black
hole binary for different black hole masses.} Shown are the binary separation
(solid), the Roche lobe sizes for the Wolf-Rayet star (dotted) and for the
black hole (dashed), the capture radius for the black hole when using the
terminal velocity (dash-dotted) or when using a simplified velocity law $v(r) =
v_\infty (1 - R_*/r) $ (long-short dashed). }

{\noindent {\bf Extended Data Figure 4. The black hole accretion rate for
different black hole mass.} The accretion rates are computed adopting the
terminal velocity (dotted) and a simplified velocity law $v(r) = v_\infty (1 -
R_*/r)$ (solid). To power the observed average luminosity of $3\times10^{38}$
erg/s,  the black hole mass must exceed $13M_\odot$ ($8M_\odot$) using the
terminal velocity (the velocity law) for a Kerr black hole, and exceed
$46M_\odot$ ($28M_\odot$) for a Schwarzschild black hole. }

{\noindent {\bf Extended Data Figure 5. Disk temperature structures for M101
ULX-1. } (a) The disk temperature profiles for M101 ULX-1 (for $P = 8.24 {\rm
days}, M_* = 19 M_\odot, R_* = 10.7 R_\odot, M_\bullet = 10/100M_\odot$) and
NGC300 X-1 (for $P = 32.4{\rm hr}, M_* = 26 M_*, R_* = 7.2R_\odot, M_\bullet =
16.9M_\odot$; Crowther et al. 2010). (b) The disk temperature at the outer edge
for different black hole mass in M101 ULX-1. }

{\noindent {\bf Extended Data Table 1. Gemini/GMOS spectroscopic observations
of M101 ULX-1.} The columns are: (1) Observation date, (2) Modified Julian Date,
(3) exposure time in seconds, (4) barycentric correction computed with {\tt
rvsao}, and (5) the corrected radial velocity as measured with HeII 4686, with
an error of 15 km/s as mainly from the uncertainties in the wavelength
calibration. }

{\noindent {\bf Extended Data Table 2. Properties of emission lines.} The
columns are: (1) emission line ID, (2) FWHM as obtained from Gaussian fit,
which equals to $2.35\sigma$, (3) equivalent width, (4) line luminosity in unit
of $10^{34}$ erg/s, and (5) equivalent width from the best WR synthetic model.
}

\newpage

\begin{figure}
\includegraphics[angle=270,origin=c,width=150mm]{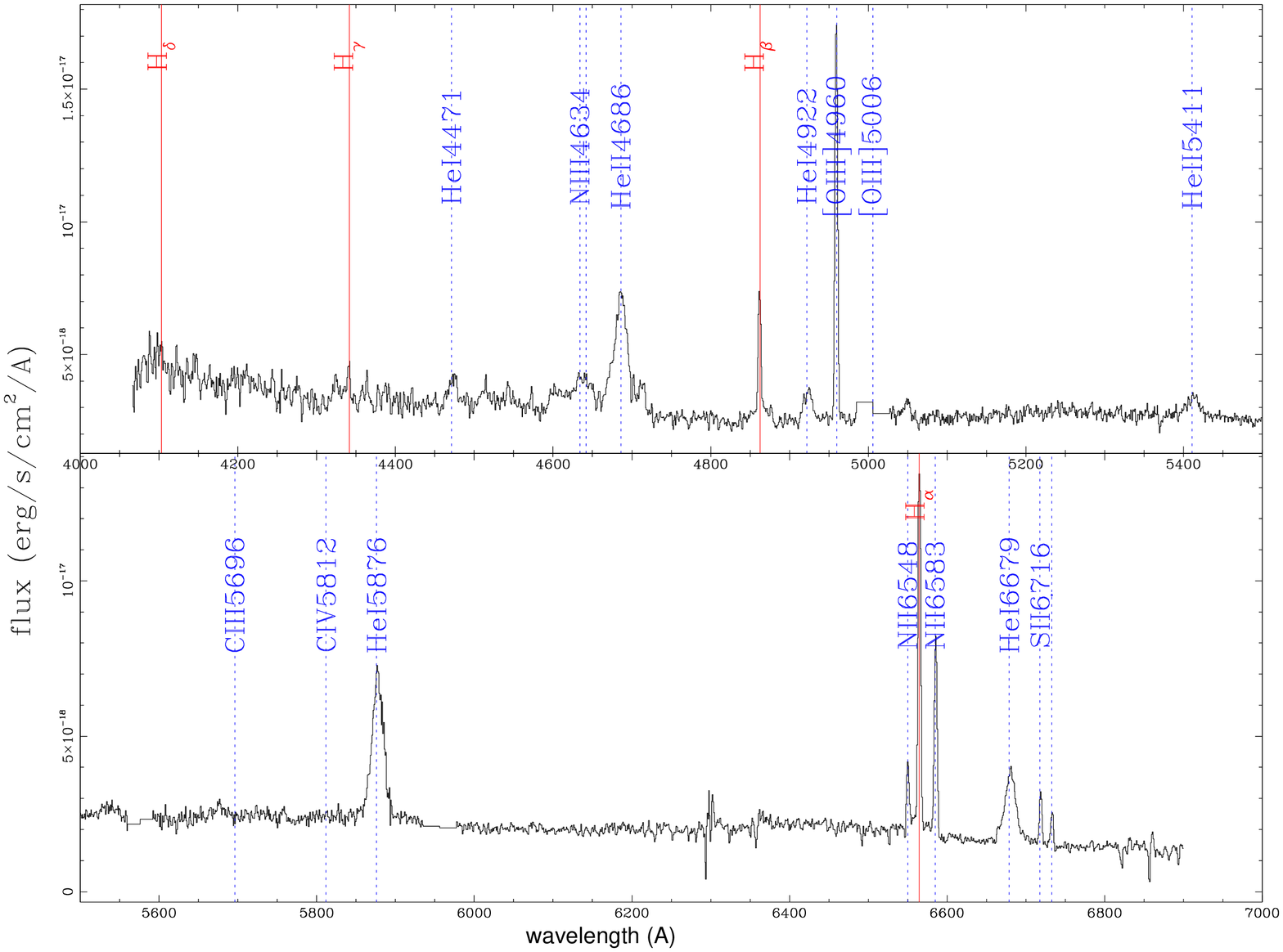}
\end{figure}

\begin{figure}
\includegraphics[width=6in,height=6in]{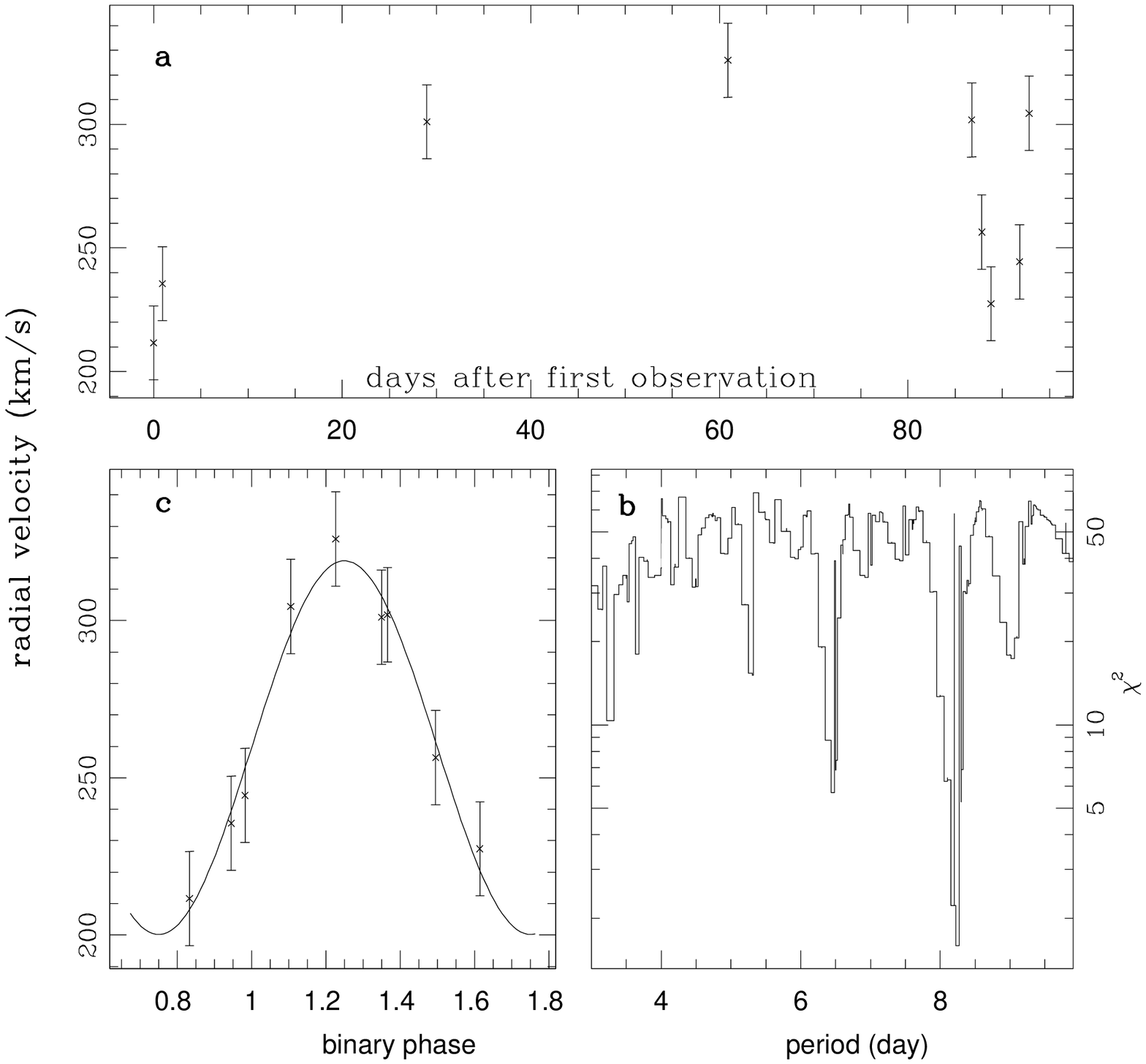}
\end{figure}

\begin{figure}
\includegraphics[width=6in,height=6in]{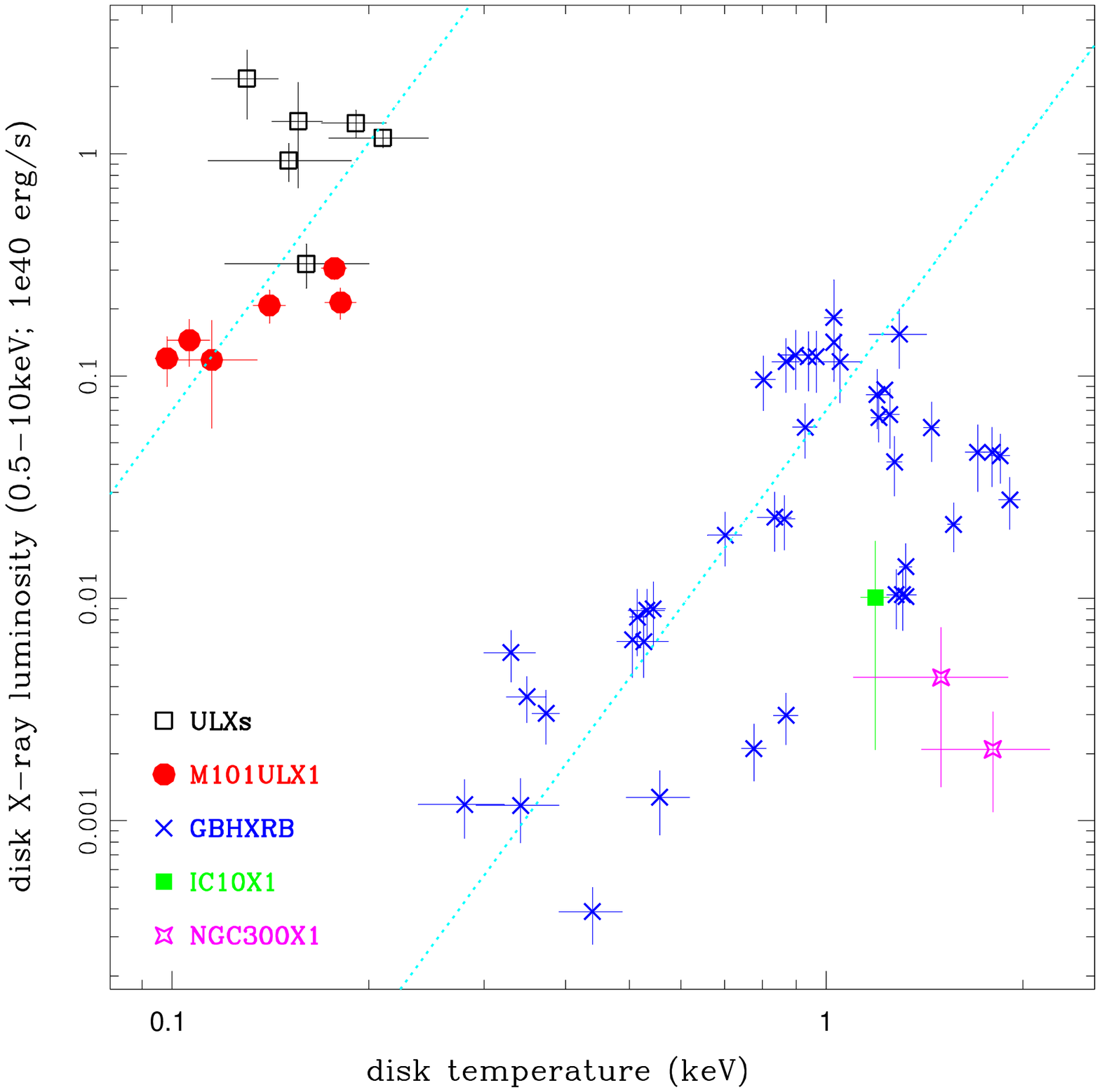}
\end{figure}

\newpage

\setcounter{page}{1}

{\noindent \large Online Methods} \\

This Online Methods part provides details about background information
for M101 ULX-1, data reduction and analysis of the Gemini/GMOS spectroscopic
observations, search for the orbital periodicity, and properties of the
Wolf-Rayet/black hole binary. It contains 5 figures
and 2 tables, and additional references.

\section{M101 ULX-1 is an outstanding IMBH candidate}

M 101 is a nearby face-on grand design spiral galaxy, a frequent target of
various observations. These include the optical monitoring observations in
search of Cepheids with the Hubble Space Telescope, yielding a distance of
6.855 Mpc\cite{Freedman01}.
%
M101 ULX-1 (CXO J140332.3+542103) is located near a spiral
arm (Extended Data Figure~1), and identified with a unique optical counterpart of $V=23.5$ mag\cite{Kong05a}.
%
At this location, the metallicity is $0.4\times$ solar according to the M101
gas-phase oxygen abundance gradient\cite{Bresolin07}.
%

This ULX has been observed intensively by X-ray missions including ROSAT, XMM
and Chandra since early 1990's, which exhibited spectral state transitions
between the low-hard state and the high-soft state reminiscent of Galactic
black hole X-ray binaries.
This ULX was once the brightest X-ray point source in M101 with a Chandra/ACIS
count rate of 0.10 count/second\cite{Liu11}, observed during the 2000 March
observation (ObsID 934).
The Chandra/XMM-Newton spectra during its outbursts\cite{Kong04,Kong05b} were
very soft and can be generally fitted with an absorbed blackbody model with
$n_H = 1-4\times10^{21}$ cm$^{-2}$ and temperatures of 50-100 eV, and the peak
0.3-7 keV luminosity reached $3\times10^{40}$ erg/s, with a bolometric
luminosity of about $10^{41}$ erg/s, suggesting an intermediate mass black
holes of a few thousand solar masses.
It was argued that it is unphysical to adopt a high neutral absorber column
density of $\ge10^{21}$ cm$^{-2}$, and fitting the spectra as blackbody plus a
{\tt diskline} component centered at 0.5 keV with $N_H$ fixed at the Galactic
value of $4\times10^{20}$ cm$^{-2}$ yielded the maximum outburst bolometric
luminosity  of $3\times10^{39}$ erg s$^{-1}$, consistent with the Eddington
luminosity of a black hole of 20-40 $M_\odot$\cite{Mukai05}.

Even at the lowered luminosities of $3\times10^{39}$ erg s$^{-1}$, the
combination of the disk luminosities and disk temperatures makes M101 ULX-1 an
outstanding IMBH candidate.
It is believed that the accretion disks for IMBHs should have larger inner
radii and consequently lower disk temperatures\cite{Miller04,Kong04,Liu08},
occupying the upper left portion in the $T_{disk} \sim L_X$ plane as shown in
Figure~3.
The position of M101 ULX-1 on this plane suggests that it is distinctly
different from the Galactic BH X-ray binaries in the lower right portion, but
belongs to the league of IMBH candidates along with some extreme ULXs above
$10^{40}$ erg/s.
The practice of placing these ULXs on this plane was questioned because
decomposing ULX spectra into DISKBB+PL is unphysical given the dominance of the
hard power-law component.  However, in the case of M101 ULX-1 the spectra are
supersoft without any hard power-law component, so its location on the
plane should reflect the accretion disk uncomplicated by Comptonization.
For comparison, we also put on this plane the other two known WR/BH
binaries\cite{Barnard08} IC 10 X-1 and NGC 300 X-1, which apparently belong to
the league of stellar mass black holes, and dynamical mass measurements have
yielded mass estimates of $20-30M_\odot$.

Combined analysis of 26 HST observations and 33 X-ray observations over 16
years\cite{Liu09} revealed two optical outbursts in addition to 5 X-ray
outbursts.  While there is no ``exact'' period for the recurring
outbursts, the outbursts occur once roughly every six months. Such outbursts
last 10-30 days, suggesting a outburst duty cycle of 10\%-15\%. Outside
outbursts, ULX-1 stays in a low-hard state with  an X-ray luminosity of
$2\times10^{37}$ erg/s\cite{Kong04,Kong05b,Mukai05,Liu09}.
%
Such behaviors is reminiscent of those of soft X-ray transients in low-mass
X-ray binaries, albeit with higher luminosities and lower disk temperatures,
but are different from the recently discovered high mass fast transients due to
clumping winds at much lower X-ray luminosities ($\sim10^{34}$ erg/s).
Detailed studies of the optical spectral energy distribution, after removal of
optical emission from the X-ray irradiated accretion disk in the outbursts,
suggest that the secondary is a Wolf-Rayet star of initially 40--60 $M_\odot$,
currently 18--20 $M_\odot$, 9--12 $R_\odot$ and about $5\times10^4$
Kelvin\cite{Liu09}.
This claim of a WR companion is supported by the presence of the He II
$\lambda4686$ emission line in the Gemini/GMOS-N spectrum taken in
2005\cite{Kuntz05}.

\section{Gemini/GMOS data reduction}

M101 ULX-1 was monitored spectroscopically from February to May in 2010 during its
expected low states under
the Gemini/GMOS-N program GN-2010A-Q49 (PI: Jifeng Liu).
Extended Data Table~1 lists the observations taken in ten nights distributed from February to
May, with a total exposure of 15.6 hours.
All exposures were taken with the $0.^{\prime\prime}75$ slit and the B600 grating tuned
for a wavelength coverage from 4000\AA\ to 6900\AA; such a slit/grating
combination will yield a spectral resolution of about 4.5\AA.
We followed standard procedures to reduce the observations and extract 1-D
spectra using the {\tt gmos} package in {\tt IRAF}.
All consecutive sub-exposures during one night were combined into one spectrum
to increase the signal-to-noise ratio, and we obtained ten spectra with
exposure times ranging from 3200 seconds to 9600 seconds (Extended Data Table~1).

For each spectrum, the wavelength solution was obtained using the Copper-Argon
arc lamp spectra taken with the same slit/grating setting right before and
after the science exposures during the same night or occasionally the night
after.
We verified the wavelength solution by comparing thus obtained wavelengths to
the intrinsic wavelengths for a dozen of strong night sky emission lines
identified in the spectra before sky subtraction, and revealed wavelength
differences with a dispersion of about 0.25\AA, or $\sim15$ km/s.
The extracted spectra were converted to flux spectra using the standard star
HZ44 taken during the night of February 15, and we scaled the spectra to have
$f_\lambda = 1.5\times10^{-18}$ erg/s/cm$^2$/\AA\ at 5500\AA\ corresponding to
F555W = 23.5 mag based on previous HST/WFPC2 observations\cite{Liu09}.
%

Figure~1  shows the flux-calibrated
sky-subtracted spectrum combined from the ten spectra.  The combined spectrum
is free of absorption lines but abundant in emission lines as identified and
listed in Extended Data Table~2.
For each emission line, we fit a Gaussian profile to derive its line width and
compute its line flux and luminosity.
Two categories of lines are present in the spectrum.
The first category is the broad helium emission lines with FWHM of up to 20\AA,
five times broader than the instrumental spectral resolution, and includes
strong HeII 4686, HeI 5876, HeI 6679, and weaker HeI 4471, HeI 4922, and HeII
5411 lines. The broad NIII 4634 emission line is also present.
The second category is the the narrow emission lines with line widths
consistent with the instrumental spectral resolution, and includes the Balmer
lines and forbidden lines such as [OIII] 4960/5006 (the latter is mostly in the
CCD gap and not listed),  [NII] 6548/6583, and [SII] 6716/6731.

The emission line properties are derived from the Gaussian line profile
fitting.  The average line properties including FWHM, Equivalent Width, and
line luminosities are measured from the combined spectrum (Extended Data Table~2).
The shifts of the line centers were also measured for individual spectra, with
the barycentric correction computed using the {\tt rvsao} package in {\tt IRAF}
as listed in Extended Data Table~1 for each spectrum.
It was found that the line shifts, after barycentric correction, are consistent
with being constant for narrow emission lines over all observations at
$230\pm15$ km/s, consistent with the radial velocity of 241 $\pm$ 2 km/s for
the face-on M101.
However, the broad helium emission lines, as measured with the strongest He II
4686 line, shifted from observation to observation between 210 km/s and 330
km/s as listed in Extended Data Table~1, with an average of 270 km/s that is significantly
different from that for nebular lines.

The properties of the nebular lines help to determine the environmental
metallicity and the neutral hydrogen column density.
$N2 \equiv {\rm [NII]} \lambda6583/H_\alpha$ can be used as an abundance
indicator\cite{Pettini04} with 12 + log(O/H) = 8.90 + 0.57 $\times N2$, albeit
with a large dispersion in log(O/H) of $\pm0.41$. Given the equivalent width of
these two lines (Extended Data Table~2), we find 12 + log(O/H) = 8.70, close to solar
metallicity (8.66). This is higher than but marginally consistent with the
value of $0.4\times$ solar according to the M101 gas-phase oxygen abundance
gradient\cite{Bresolin07}
given the location of ULX-1. The observed Balmer line flux ratios can be used
to infer the dust extinction between the nebula and the observer.  In the
nebular emission around ULX-1, the intrinsic ratio $H_\alpha/H_\beta$ is 2.74 in
case B for a thermal temperature of T=20,000K\cite{Osterbrock89}.  Assuming
E(B-V) = 0.1 mag, then $A_{6564} = 0.250$mag, $A_{4863} = 0.360$mag, $\Delta A$
= 0.11 mag, $\Delta H_\alpha/\Delta H_\beta = 1.1$, and reddened
$H_\alpha/H_\beta$ $\sim3$. The observed $H_\alpha/H_\beta$ is 2.85, suggesting that
the extinction is low, and using the Galactic value is reasonable.

\section{ULX-1 is a Wolf-Rayet binary of the WN subclass}

The broad helium emission lines in the newly obtained Gemini/GMOS spectrum are
typical of an extremely hot, hydrogen depleted Wolf-Rayet star.
Accretion disks around a compact object can also give rise to broad helium
emission lines, but broad Balmer line is expected to be present and much
stronger than the helium lines.
Indeed, broad $H_\beta$ emission lines are present in two ULXs with optical
spectra (4000-5400\AA), NGC1313 X-2\cite{Roberts11} and NGC 5408
X-1\cite{Cseh11}, and are stronger than the He II 4686 emission line.
In the ULX-1 spectrum (Figure~1), while the Balmer emission lines are present,
they are narrow emission lines like forbidden lines, and should come from the
surrounding nebulae, as evidenced by their nearly constant line shifts from
observation to observation, in distinct contrast to helium lines with line
shift difference of $\pm60$ km/s.

The sub-type of this Wolf-Rayet star can be determined from the presence or
absence of line species in the spectrum\cite{Crowther07}.
%
There are two main types of Wolf-Rayet stars, WN stars with $R\sim5-12R_\odot$
revealing H-burning products and subsequently more compact WC stars with
$R\sim2-3R_\odot$ revealing He-burning products.
WC stars are dominant by carbon lines (such as CIII 4650, CIII 5696 and CIV
5812) that are stronger than helium lines, but none of the
carbon lines are present in the ULX-1 spectrum.
WN stars from WN4 to WN8 show\cite{Hamann93} increasing absolute magnitudes $M_V$ from -3.5
mag to -6 mag, increasing mass loss rates from $10^{-5}$ to $10^{-4}$
$M_\odot/yr$, decreasing effective temperatures from 80 kK to 45 kK,
hence increasing fraction of HeI atoms relative to HeII ions.
Comparing the observed spectrum to the spectral atlas of WN stars\cite{Crowther07,Crowther06},
we estimate a late-type WN8 star.  A WN8
subtype is also inferred based on the HeI 5876/He I 5411 equivalent width ratio\cite{Smith96}.
%
Such a subtype is roughly consistent with its absolute magnitude of $M_V$ = -
5.9 mag (after extinction correction using Galactic E(B-V) = 0.1 mag and $R_V$
= 3.1), and the effective temperature of about 50 kK derived from its
broad-band spectral energy distribution\cite{Liu09}.
%

\section{Physical parameters for the Wolf-Rayet star}

As for the case of NGC 300 X-1\cite{Crowther10},
we have calculated
synthetic models using the
line-blanketed, non-local
thermodynamic equilibrium model atmosphere code\cite{Hillier98}.
To select the best physical parameters of the WR star, we compare the model
equivalent width with observed values for the six helium emission lines and
minimize the quantity $\Delta^2 = \sum_i ({\rm EW}-{\rm EW}_i)^2$.
In all model calculations, elemental abundances are set to 40 percent of the
solar value for the metallicity of $0.4Z_\odot$ at the location of ULX-1.
We vary the stellar radius $R_*$ between 4 to $20R_\odot$, stellar mass $M_*$ between
$5-35M_\odot$, stellar luminosity $L_*$ between $5-100\times10^4L_\odot$, the outer
radius for line-forming region $R_{\rm MAX}$ up to $40R_\odot$, the terminal
velocity $v_\infty$ between $400-2000$ km/s, and the stellar wind mass loss rate
$\dot{M}_*$ between $5-100\times10^{-6} M_\odot/yr$.

We have run $\sim5000$ models with the combination of stellar mass, radius and
luminosity determined by the stellar evolution tracks\cite{Girardi02} of
$Z=0.4Z_\odot$ for all possible WN stars,
and another $\sim5000$ models with ``fake'' stars whose mass, radius and
luminosity are completely independent of each other.
After a total of $\sim10000$ model evaluations, a best fitting model is found
with $R_* = 10.7 R_\odot$, $M_* = 17.5M_\odot$, $L_* = 5.4\times10^5L_\odot$,
$v_\infty = 1300$ km/s, $R_{\rm MAX} = 22 R_\odot$, and $\dot{M}_* =
2.0\times10^{-5} M_\odot/yr$.
The model reproduces the helium emission lines extremely well (Extended Data Table~2), with
an average difference of $|\Delta|$ = 0.6\AA.  In comparison, the majority of
models and all models with ``fake'' stellar parameters are much worse-fitting
with $\Delta^2 >> 10$ (Extended Data Figure~2).  Based on the $\Delta^2$ distribution, our
model evaluations picked up the stellar parameters effectively, and we
estimate, with equivalently $\Delta\chi^2 = 1$,
the errors  to be $\dot{M}_* = 2\pm 0.5 \times 10^{-5} M_\odot/yr$, and
$v_\infty = 1300 \pm 100$ km/s.
Note that, if we adopt a solar metalicity, as allowed by the abundance
indicator $N2 \equiv {\rm [NII]} \lambda6583/H_\alpha$, the best model will
change to  $R_* = 11.1 R_\odot$, $M_* = 17.5M_\odot$, $L_* =
4.9\times10^5L_\odot$, $v_\infty = 1700$ km/s, and $\dot{M}_* =
2.4\times10^{-5} M_\odot/yr$.
This is consistent with the $0.4Z_\odot$ results within the errors except for a
significantly higher terminal velocity.

The stellar parameters of this best model belong to a ``real'' WN star from the
stellar evolution tracks, with an effective temperature of 48 kK, an initial
mass of $42M_\odot$, an age of about 5 Myrs, and a remaining lifetime of about
0.3 Myrs before it loses another $\sim6M_\odot$ and  collapses into a black
hole of $\sim12M_\odot$. This model is actually one of the best models derived
from studies of the optical spectral energy distribution\cite{Liu09}.
Comparing to the physical properties of WR stars in the Milky
Way\cite{Crowther07},
we find that $T_*, L_*, \dot{M}_*$, and $v_\infty$ are consistent with those
for a WN7/WN8 star.  The absolute magnitude $M_V$ for ULX-1 ($M_V = -5.9$ mag
after extinction correction) is brighter by 0.5 mag, fully within the spread of
absolute magnitudes for WN subtypes.

The mass of the WR star can be more reliably estimated with the the empirical
mass-luminosity relation\cite{Schaerer92,Crowther07} as done for NGC 300
X-1\cite{Crowther10}.
%
In our case, $L_* = 5.4\times10^5L_\odot$, this corresponds to a WR mass of $19
M_\odot$, quite consistent with the mass for the best model.
The luminosity derived for solar metalicity will correspond to a WR mass of
$18M_\odot$.
Hereafter we will use $19M_\odot$ for the WR mass, with an estimated formal
error of $1M_\odot$ to roughly reflect the difference between the model value
and the empirical value.
Given the stellar mass and radius of $10.7R_\odot$, we can obtain the orbital
period\cite{FKR2002} as $P = \sqrt{\rho/110} {\rm hr} \simeq 72 {\rm hr}$ if
the WR star is filling its Roche lobe.
The true orbital period will be longer than 72 hrs if the WR star is only
filling part of its Roche lobe.

\section{Search for orbital periodicity}

The radial velocity changes between 210 km/s and 330 km/s as measured by the
HeII 4686 emission line should reflect the orbital motion of the WR star.
While broad HeII 4686 emission line can be produced from the X-ray heated
accretion disk in some ULXs with rather high X-ray luminosities (e.g., in NGC
1313 X-2 with $\sim10^{40}$ erg/s\cite{Roberts11}), this should not be the case
for M101 ULX-1 because its X-ray luminosities during the Gemini/GMOS
observations were three orders of magnitude lower, and the disk heating effects
are insignificant even in its outburst based on the optical
studies\cite{Liu09}.  In addition, the line ratios for the heated accretion
disk are different from the line ratios for the WR star because the emission
line forming regions and temperature structures are quite different, yet the
observed line ratios can be well reproduced by the WR star.

In order to search for the orbital periodicity, we assume a circular orbit and
fit a since curve $v_r = v_0 + K \sin[2\pi(t-t_1)/P+\phi]$ to nine
barycenter-corrected radial velocities; the radial velocity for March 17th was
dropped from the analysis because the spectrum had a very low signal-to-noise
ratio.  The four parameters are the radial velocity of the binary mass center
$v_0$, the radial velocity semi-amplitude $K$, the orbital period $P$, and
phase $\phi$ at the first observation.
The search is carried out by minimizing $\chi^2$ defined as $\chi^2 =
\sum_{i=1}^{10} [v_r(t_i)-v_{r,i}]^2/\sigma_{v_ri}^2$. The radial velocity
errors $\sigma_{v_ri}$ are taken as the wavelength calibration error of
0.25\AA, or 15 km/s.
The five radial velocity measurements from May 13th to May 19th suggest a
period no longer than 10 days (Figure~2).
The amoeba technique is used for $\chi^2$ minimization, using initial guesses
taken from the parameter grids with $P$ from 3 to 10 days in step of 0.01 days,
$K$ from 20 to 150 km/s in step of 5 km/s, and $\phi$ from $0^\circ$ to
$360^\circ$ in step of $10^\circ$.
The best solution is found at the minimum $\chi^2 = 1.6$, for which the best
period $P=8.24 \pm 0.1$ days and the best radial velocity semi-amplitude $K =
61 \pm 5$ km/s, with the 68.3\% error determined with $\Delta\chi^2 = 1$.
The fact that the radial velocity curve can be fitted with a sine curve
suggests that the orbital eccentricity is small.

Given $P$ and $K$, the mass function for M101 ULX-1 can be computed as
$f(M_*,M_\bullet,i) = {P K^3 \over 2\pi G} =  {M_\bullet^3 \over (M_\bullet +
M_*)^2} \sin^3i = 0.178 M_\odot$. This sets an absolute lower limit for the
mass of the primary. In the case of ULX-1, more information can be extracted
because we already know $M_* = 19 M_\odot$.  Given the equation ${M_\bullet^3
\over (M_\bullet + M_*)^2} \sin^3i = 0.178 M_\odot$, the primary mass will
increase monotonically when the inclination angle decreases, i.e., changing
from edge-on ($i=90^\circ$) toward face-on ($i=0^\circ$).
Thus the minimum mass for the primary can be obtained when $i=90^\circ$, which
is $M_\bullet = 4.6 M_\odot$ after solving the equation $ {M_\bullet^3 \over
(M_\bullet + M_*)^2} = 0.178 M_\odot $. The minimum mass will be $M_\bullet =
4.4 M_\odot$ if we use $M_* = 17.5 M_\odot$. Such a compact primary can only be
a black hole. This is thus the dynamical evidence for a black hole in a ULX.


\section{The Wolf-Rayet/black hole binary properties}

This section duplicates some text from the main article, but with additional
technical details.

M101 ULX-1 is thus a Wolf-Rayet/black hole binary, only the third discovered so
far after IC 10 X-1 and NGC300 X-1.
The binary separation can be computed with Kepler's Law $a^3 =
{G(M_*+M_\bullet) \over 4\pi^2} P^2$, which increases monotonically for
increasing black hole mass, starting from $a = 50R_\odot$ for $M_\bullet =
4.6M_\odot$ to $a = 75R_\odot$ for $M_\bullet = 60M_\odot$ (Extended Data Figure~3).
The Roche lobe size for the secondary can be computed with $R_{cr} = a\cdot
f(q) = a \cdot 0.49 q^{2/3}/[0.6 q^{2/3} + ln(1+q^{1/3})]$ with $q =
M_*/M_\bullet$, and the Roche lobe size for the black hole can be computed with
the same formula but with different $q = M_\bullet/M_*$.
As shown in Extended Data Figure~3, the Roche lobe size for the black hole increases with
the increasing black hole mass, but the Roche lobe size for the secondary does
not change much, from $R_{cr,*} = 25R_\odot$ for $M_\bullet=4.6M_\odot$ to
$R_{cr,*} = 23R_\odot$ for $M_\bullet=10M_\odot$, and to $R_{cr,*} = 22R_\odot$
for $M_\bullet=20M_\odot$.

Regardless of the black hole mass, the secondary is filling only half of its
Roche lobe by radius, and the black hole must be accreting from the Wolf-Rayet
star winds. Since the black hole is at least $50R_\odot$ away from the WR star,
the stellar wind must have reached close to its terminal velocity.  The capture
radius for the wind accretion can be computed as $r_{acc} = {2GM_\bullet \over
v_\infty^2}$, and the accretion rate can be computed as $\dot{M_\bullet} = {\pi
r_{acc}^2 \over 4\pi a^2} \dot{M_*}$.  Given that the average luminosity for
M101 ULX-1 is about $3\times10^{38}$ erg/s, the required accretion rate is
$\dot{M_\bullet} = L / \eta c^2 \simeq {L_{38} \over \eta} 2\times10^{-9}
M_\odot/yr = {1 \over \eta} 6 \times10^{-9} M_\odot/yr$.  To capture this much
stellar wind matter, as shown in Extended Data Figure~4, the black hole mass must be
greater than $46M_\odot$ for $\eta=0.06$ in the case of a non-spinning
Schwarzschild black hole, and greater than $13M_\odot$ for $\eta=0.42$ in the
case of a maximally spinning Kerr black hole.  If we use the velocity law $v(r)
= v_\infty (1 - R_*/r)^\beta$ with $\beta=1$ for the inner
wind\cite{Hillier98},
then the black hole mass must be greater than $28M_\odot$ for $\eta=0.06$ in
the case of a non-spinning Schwarzschild black hole, and greater than
$8M_\odot$ for $\eta=0.42$ in the case of a  maximally spinning Kerr black
hole.
If we adopt a typical $\eta$ value of 0.1, the required accretion rate
corresponds to $M_\bullet > 24 M_\odot$ (and $i<17^\circ$) for a wind velocity
of $v \simeq 1100$ km s$^{-1}$, and corresponds to $M_\bullet > 32 M_\odot$
(and $i<14^\circ$) for the terminal velocity.
The accretion rate argument thus requires a black hole of $>8-46M_\odot$,
likely a black hole of $20-30M_\odot$ similar to IC 10 X-1 and NGC 300 X-1.

The recurring X-ray/optical outbursts dictates the presence of an accretion
disk prone to instability, and the disk formation under stellar wind accretion
places stringent constraints on the binary system. To explore why the number of
Galactic X-ray stars is so small, it has been shown\cite{Illarionov75} that in
the case of accretion of stellar wind matter in a detached binary system the
specific angular momentum of the matter captured by the compact object is
typically small. Therefore, usually no accretion disk is formed around the
compact object.  Consequently, very special conditions are required for a black
hole in a detached binary system to be a strong X-ray source. A disk may form
if the specific angular momentum of accreting matter, $Q_{acc} = {1 \over 4}
{2\pi \over P} r^2_{acc}$, exceeds the specific angular momentum of the
particle at the innermost stable circular orbit, $Q_{ISCO} = \sqrt{3} r_g c =
\sqrt{3} {2GM_\bullet \over c^2} c$. This is usually expressed as $P < 4.8
{M_\bullet/M_\odot \over v_{1000}^4} \delta^2$ hr, where $\delta\sim1$ is a
dimensionless parameter\cite{Ergma98,Prestwich07}.  Given $P = 8.24 \pm 0.1 $
days and $v_\infty = 1300 \pm 100$ km s$^{-1}$ for M101 ULX-1, the black hole
mass is required to be $M_\bullet > 80 M_\odot$, corresponding to $i = 9^\circ$
(i.e., nearly face-on). If the wind velocity from the velocity model of the
inner wind\cite{Hillier98} is adopted, then the black hole mass is required to
be $M_\bullet > 48 M_\odot$, corresponding to $i = 11^\circ$.

To investigate the possible presence of partial ionization zone, we need to
compute the temperature structure $T_d(r)$ for the accretion disk, especially
for the outer disk.
Following the procedures designed for an X-ray irradiated black hole binary model for ULXs\cite{Liu12},
we compute the disk temperature structure for a standard accretion disk
with the $\alpha$ prescription\cite{Shakura73} plus X-ray irradiation\cite{FKR2002}.
%
As shown in Extended Data Figure~5, regardless of the black hole mass for M101 ULX-1, its
outer disk temperature is as low as 4000K in the low-hard state due to
its large separation and large disk, and the helium partial ionization zone at
about 15000K is bound to exists unless the black hole mass is lower than 5.5$M_\odot$.
In comparison, the disk temperature for NGC 300 X-1, with an orbital period of
32.8 hrs and its WN5 star ($M_* = 26 M_\odot, R_* = 7.2R_\odot$) filling its
Roche lobe\cite{Crowther10}, never drops below 20000K due to its small separation and small
disk, and there is no helium partial ionization zone in the disk.
This explains naturally why NGC 300 X-1 and similar IC 10 X-1 exhibit steady
X-ray radiation despite the apparent variations due to orbital modulation under
the edge-on viewing geometry.

The existence of an accretion disk in M101 ULX-1 is also supported by the
observed spectral state changes, which resemble those for Galactic black hole
binaries\cite{rem06,McClintock06} that are believed to reflect changes in the properties
of their accretion disks\cite{Esin97}.
%
During its outbursts, M101 ULX-1 exhibits an X-ray
spectrum\cite{Kong04,Mukai05} that can be classified as a thermal dominant
state (albeit with exceptionally low disk temperatures), a well-defined
spectral state that corresponds to a standard thin accretion disk at about 10\%
of its Eddington luminosity.
Quantitative studies\cite{Steiner09} show that when the luminosity exceeds 30\%
of the Eddington limit, the emission changes such that the X-ray spectrum
includes a steep power-law with a significant hard component above 2 keV. The
presence of such a hard component is not seen in the X-ray spectra of M101
ULX-1.  Given its bolometric luminosity of $3\times10^{39}$ erg s$^{-1}$ in the
thermal dominant state at less than 30\% of its Eddington limit, we infer that
the black hole mass is above $80 M_\odot$. If this is true, the inferred black
hole mass of M101 ULX-1 may challenge the expectations of current black hole
formation theories.
The most massive black holes that can be produced for solar metallicity are
about $15M_\odot$, and about $20M_\odot$ ($25M_\odot$, $30M_\odot$) for
$0.6\times$ ($0.4\times$, $0.3\times$) solar metallicity due to reduced stellar
winds and hence reduced mass loss in the final stages before stellar
collapse\cite{Belczynski10}.

\newpage

\begin{figure}
\includegraphics[width=75mm]{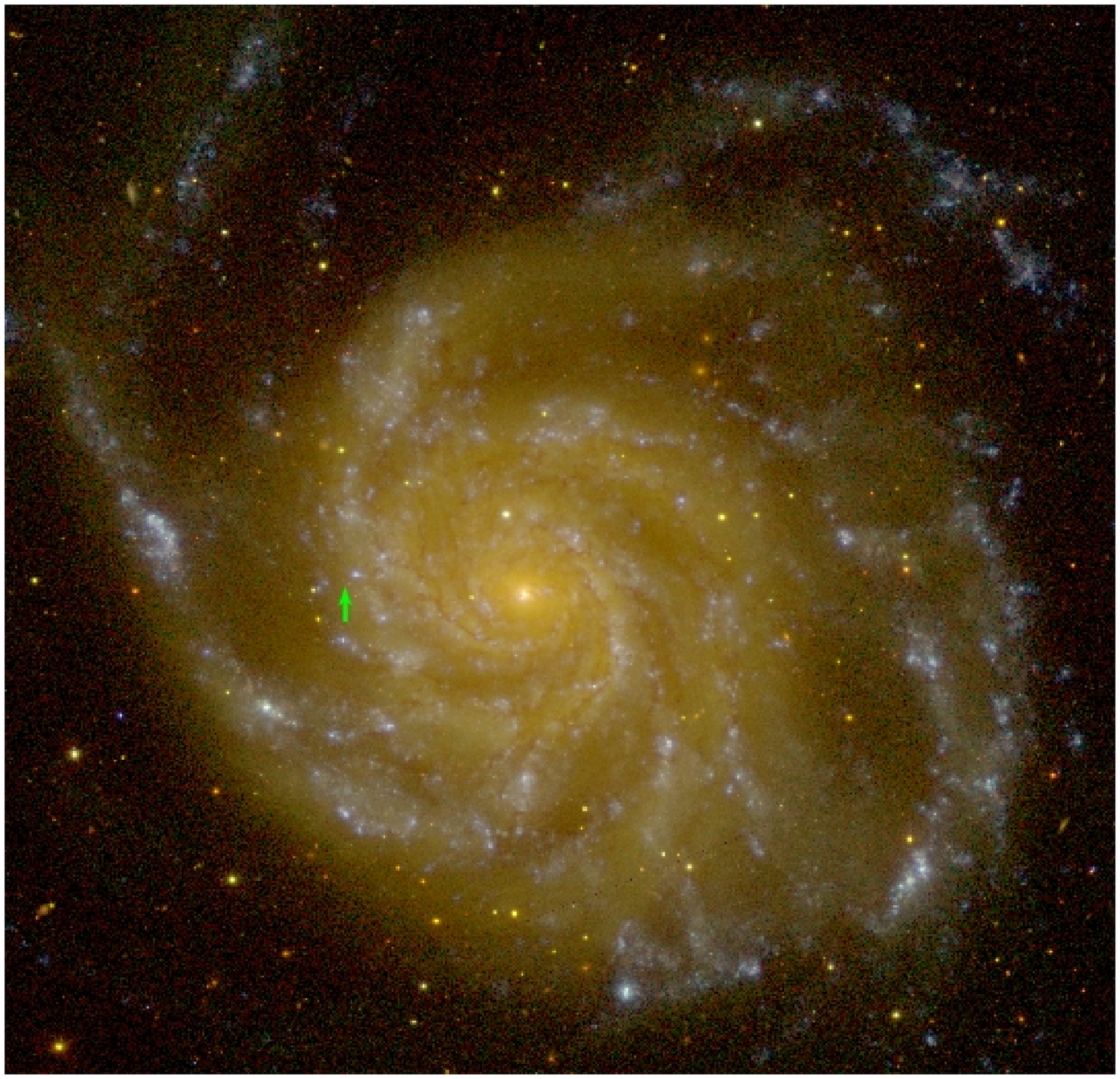}
\includegraphics[width=75mm]{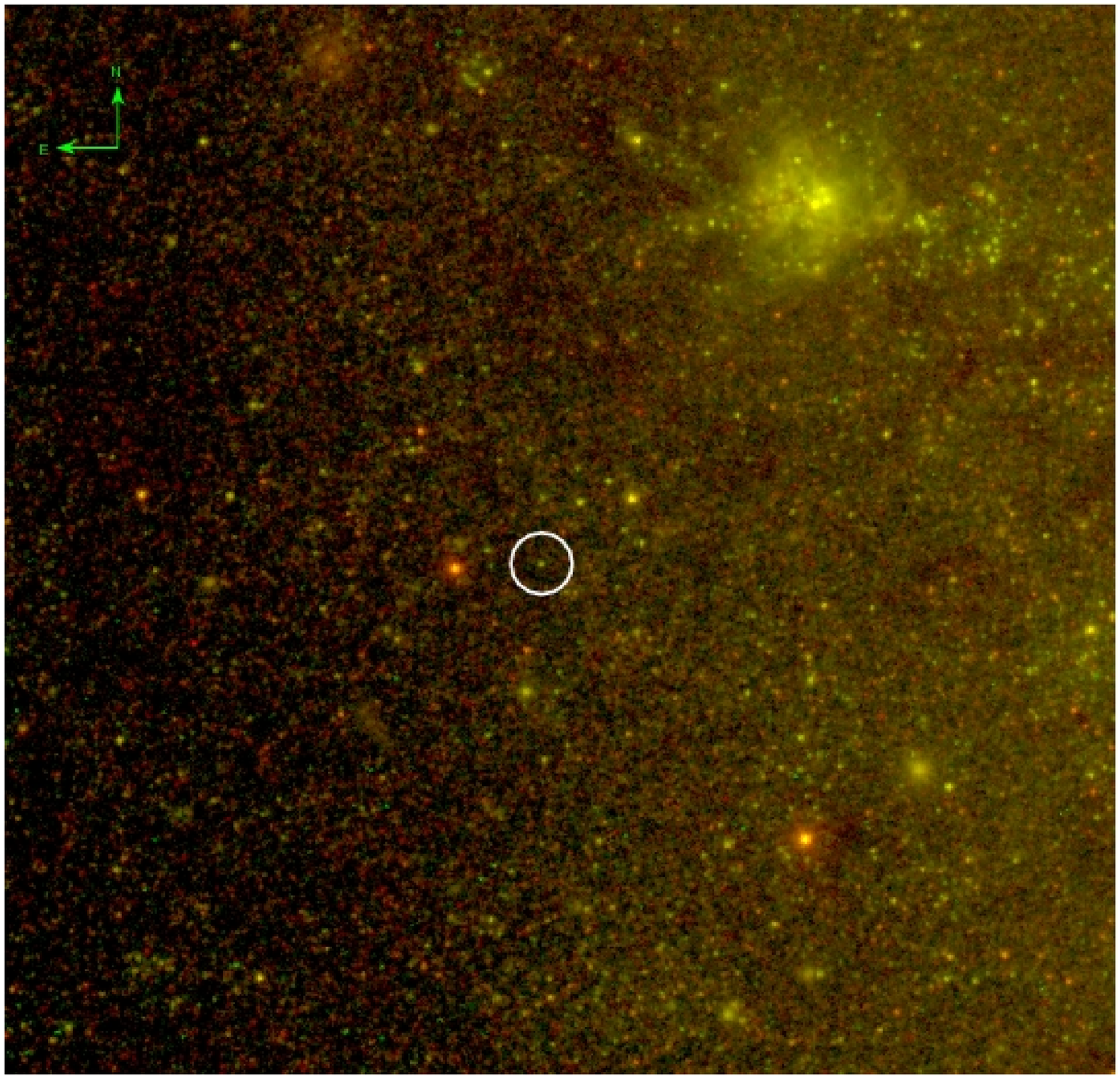}
\caption{{\bf M101 ULX-1 as observed in the optical.} (a) M101 ULX-1 is located
on a spiral arm of the face-on grand-design spiral galaxy M101, as indicated by
the arrow. The color image of M101 is composed of GALEX NUV, SDSS g, and 2MASS
J images.  (b) ULX-1 is identified as a blue object with V=23.5 mag at the
center of the $1^{\prime\prime}$ circle on the HST image. The color image is
composed of ACS/WFC F435W, F555W and F814W images. }

\end{figure}

\begin{figure}


\begin{tabular}{cc}
\parbox[b]{58mm}{\psfig{file=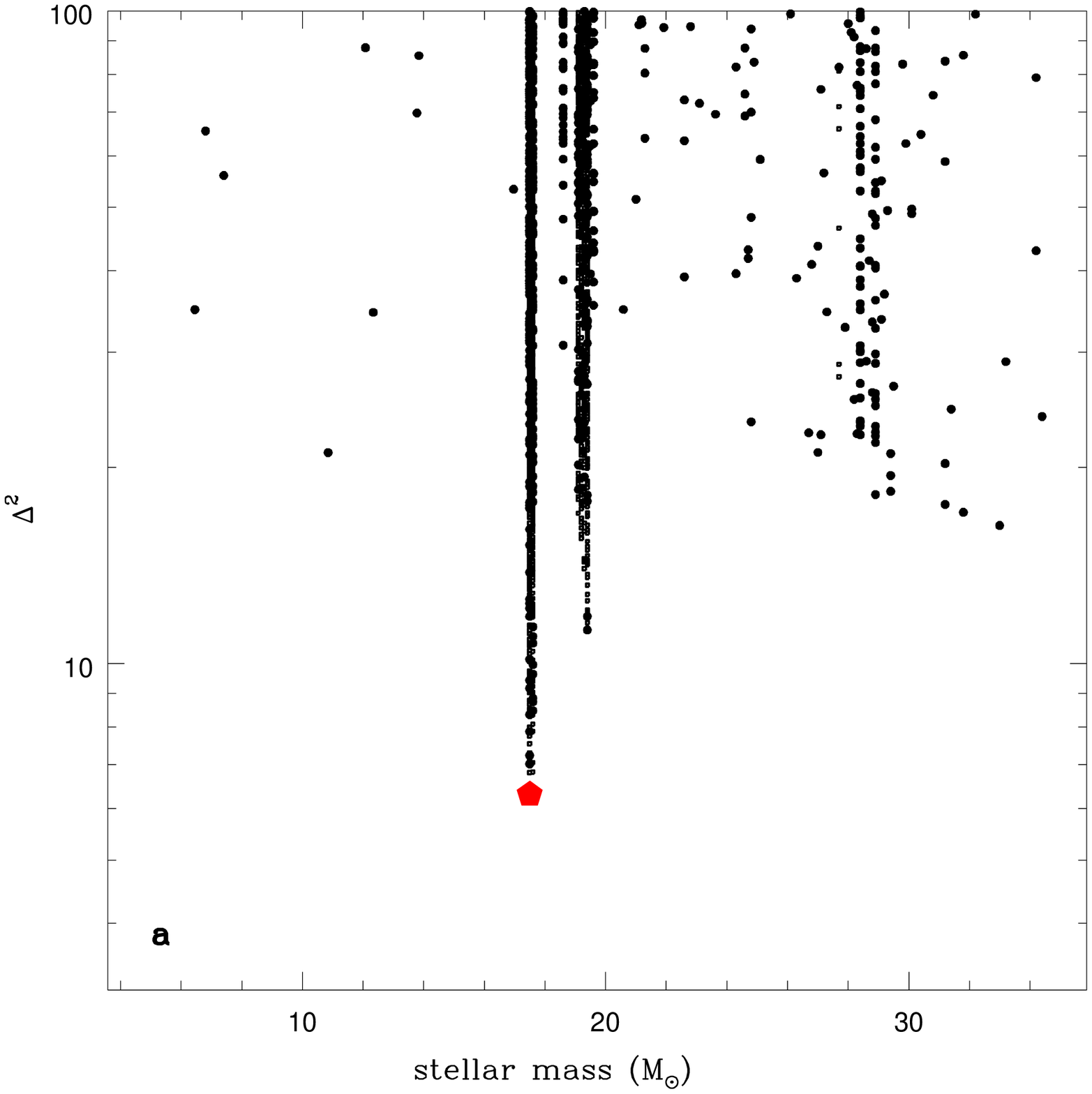,width=73mm} }\
\parbox[b]{58mm}{\psfig{file=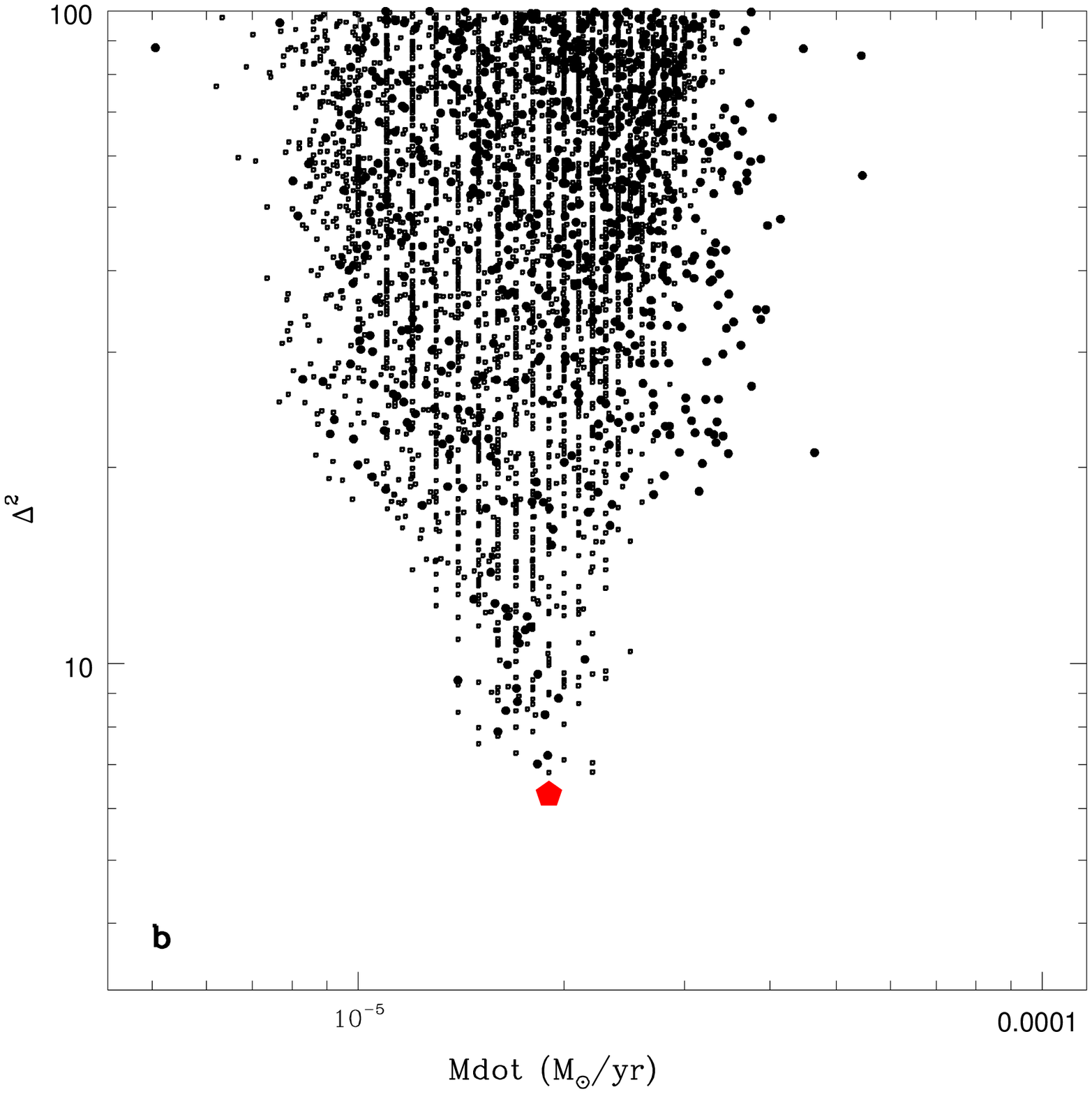,width=73mm} }\\
\parbox[b]{58mm}{\psfig{file=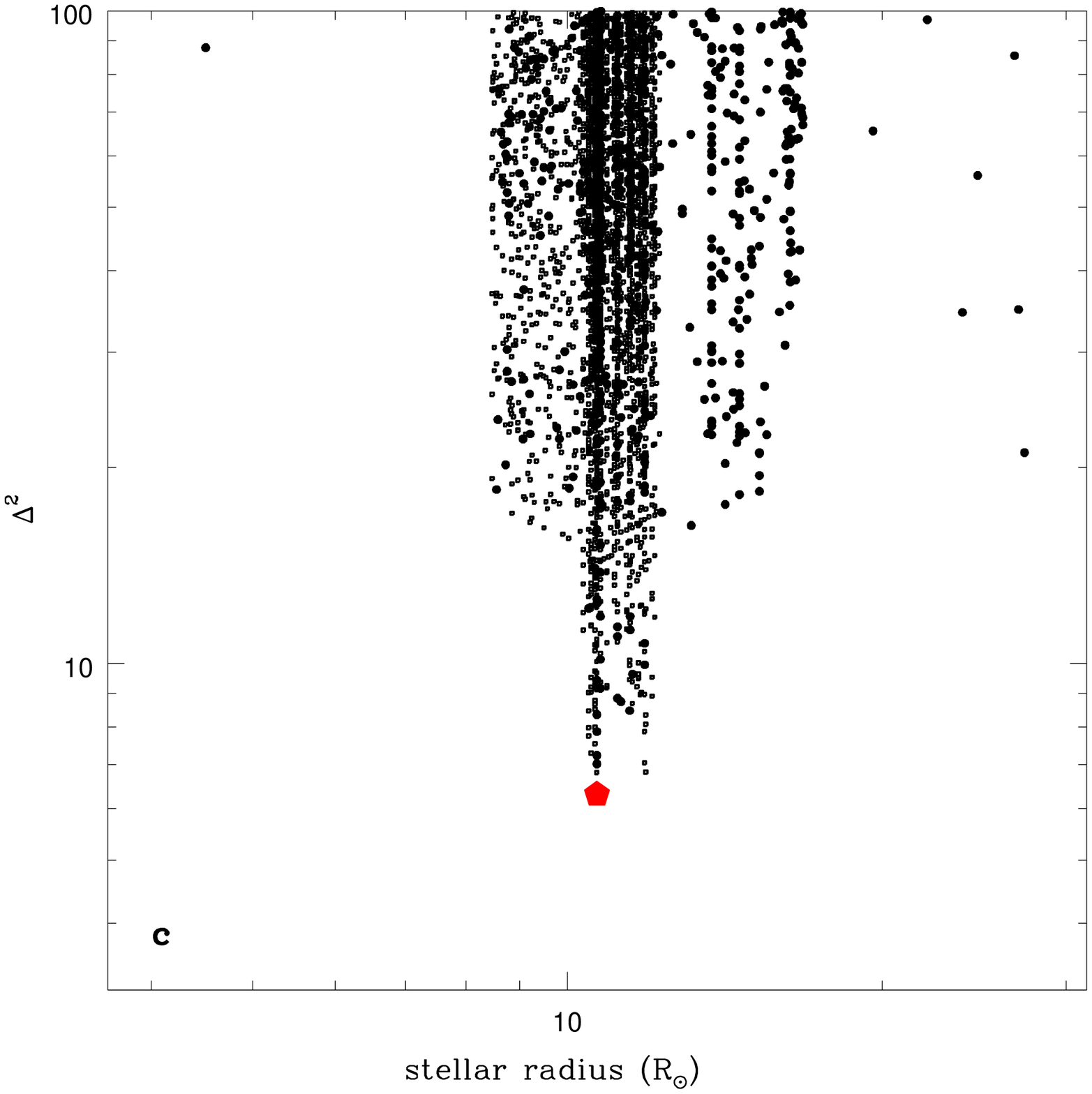,width=73mm} }\
\parbox[b]{58mm}{\psfig{file=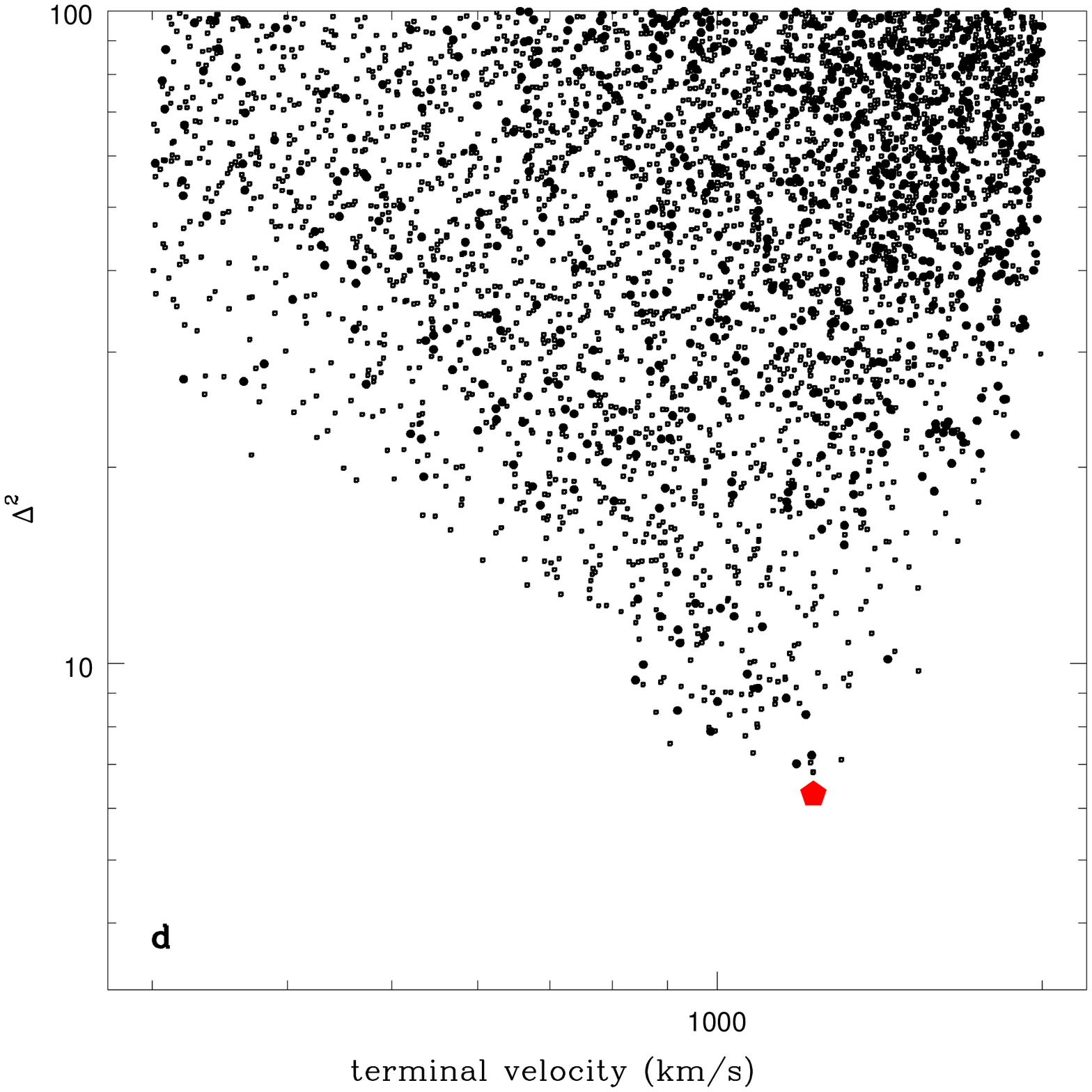,width=73mm} }\\
\end{tabular}
\caption{{\bf Physical properties of the WR secondary from spectral line
modeling.}  Distributions of computed $\Delta^2$ as a function of (a) stellar
masses, (b) stellar mass loss rate, (c) stellar radii, and (d) terminal
velocity. Here $\Delta^2 = \sum_i ({\rm EW} - {\rm EW}_i)^2$ computes the
difference between observed and synthetic equivalent widths for six broad
helium lines present in the Gemini/GMOS spectrum. We have computed synthetic
spectra for a group of 5000 real stars from the evolution tracks (as shown by
the thick stripes in the mass plot and the radius plot) and for another group
of "fake" stars with continuous distributions in mass, radius and luminosity.
The best model is labeled by a filled pentagon in all panels.  }

\end{figure}

\begin{figure}

\includegraphics[width=150mm]{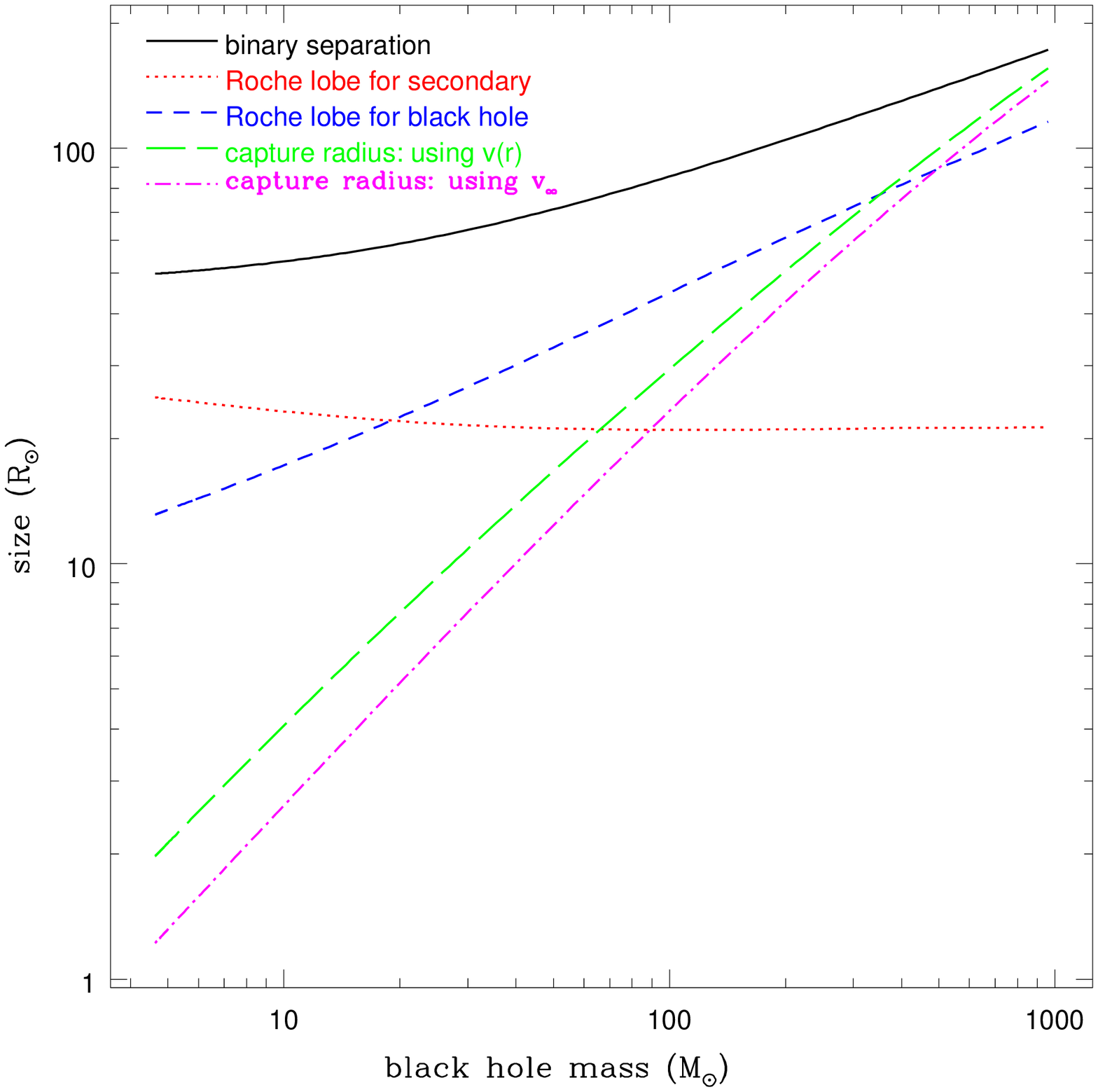}
\caption{{\bf Properties for the Wolf-Rayet/black hole binary for different
black hole masses.} Properties for the Wolf-Rayet/black hole binary for
different black hole masses. Shown are the binary separation (solid), the Roche
lobe sizes for the Wolf-Rayet star (dotted) and for the black hole (dashed),
the capture radius for the black hole when using the terminal velocity
(dash-dotted) or when using a simplified velocity law $v(r) = v_\infty (1 -
R_*/r) $ (long-short dashed). }

\end{figure}

\begin{figure}

\includegraphics[width=150mm]{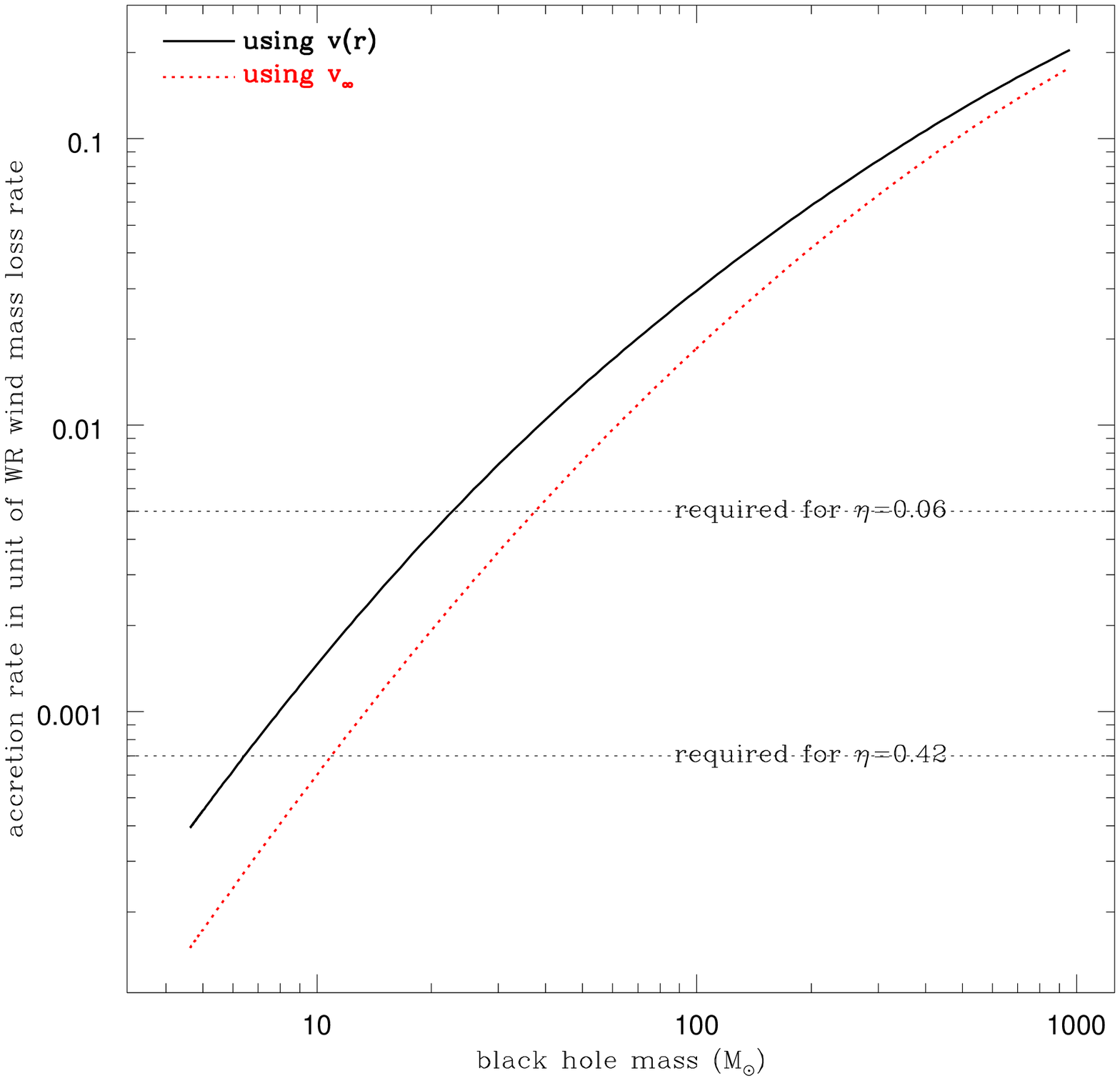}
\caption{{\bf The black hole accretion rate for different black hole mass.} The
black hole accretion rate for different black hole mass if adopting the
terminal velocity (dotted) or a simplified velocity law $v(r) = v_\infty (1 -
R_*/r)$ (solid). To power the observed average luminosity of $3\times10^{38}$
erg/s,  the black hole mass must exceed $13M_\odot$ ($8M_\odot$) using the
terminal velocity (the velocity law) for a Kerr black hole, and exceed
$46M_\odot$ ($28M_\odot$) for a Schwarzschild black hole. }

\end{figure}

\begin{figure}

\includegraphics[width=73mm]{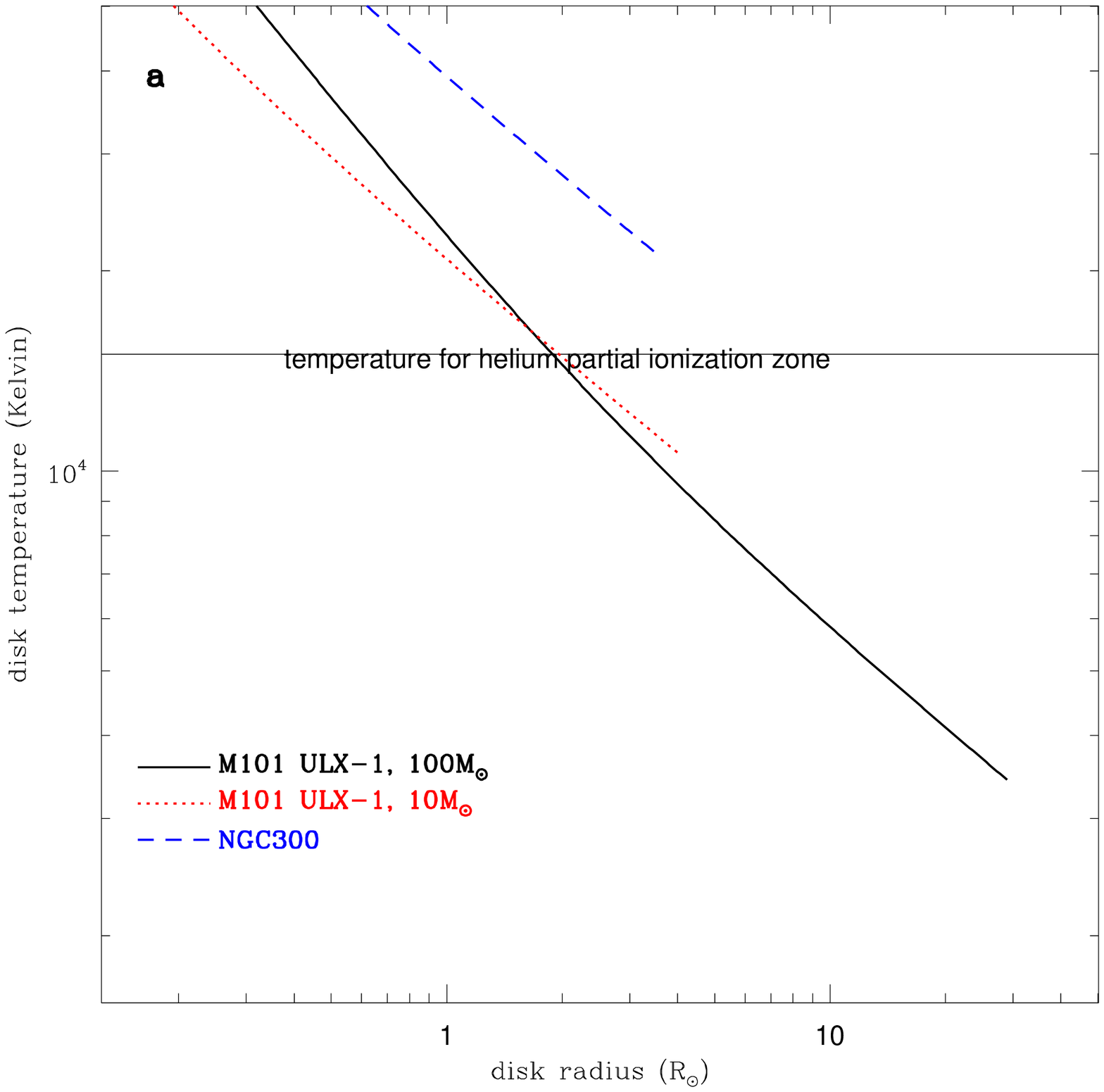}
\includegraphics[width=73mm]{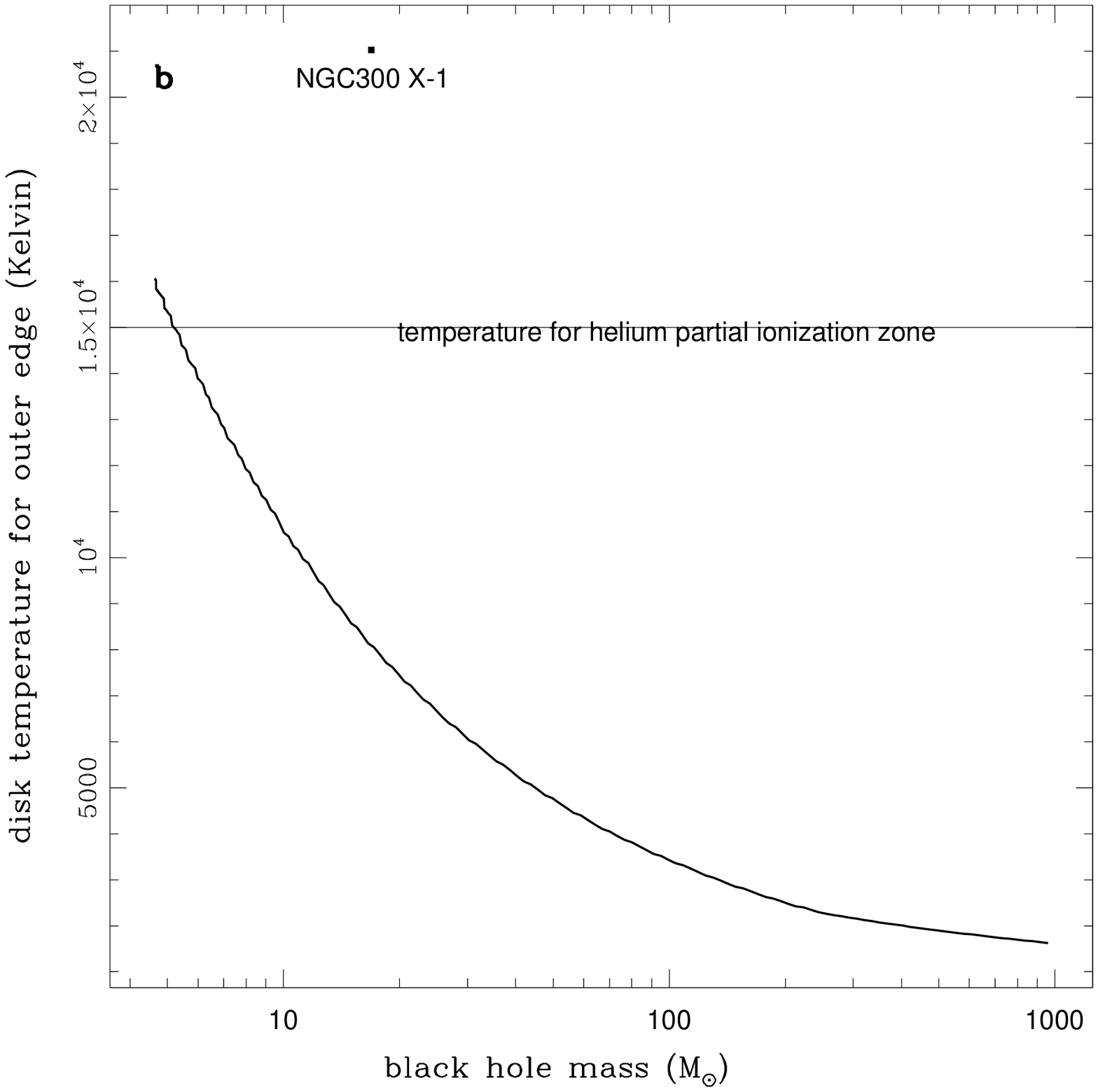}
\caption{{\bf Disk temperature structures for M101 ULX-1. } (a) The disk
temperature profiles for M101 ULX-1 (for $P = 8.24 {\rm days}, M_* = 19
M_\odot, R_* = 10.7 R_\odot, M_\bullet = 10/100M_\odot$) and NGC300 X-1 (for $P
= 32.4{\rm hr}, M_* = 26 M_*, R_* = 7.2R_\odot, M_\bullet = 16.9M_\odot$;
Crowther et al. 2010). (b) The disk temperature at the outer edge for different
black hole mass in M101 ULX-1. }

\end{figure}





\begin{table}
\begin{center}
  \begin{threeparttable}
    \caption{Gemini/GMOS spectroscopic observations of M101 ULX-1}
     \begin{tabular}{lllll}
      \toprule
      \toprule
        OBSDATE & MJD & exposure & bary.  & velocity \\
         &     & (second) &   (km/s)    &  (km) \\
      \midrule
        2010-02-15 & 55242.58343 & 3200 & 7.4 & 212 \\
        2010-02-16 & 55243.50615 & 3200 & 7.3 & 236 \\
        2010-03-16 & 55271.54390 & 3200 & 0.1 & 301 \\
        2010-03-17 & 55272.54564 & 3200 & -0.2 & --- \\
        2010-04-17 & 55303.47547 & 4800 & -7.7 & 326 \\
        2010-05-13 & 55329.33126 & 6400 & -12.2 & 302 \\
        2010-05-14 & 55330.39682 & 6400 & -12.4 & 256 \\
        2010-05-15 & 55331.37803 & 6400 & -12.5 & 227 \\
        2010-05-18 & 55334.41410 & 9600 & -13.0 & 244 \\
        2010-05-19 & 55335.42391 & 9600 & -13.1 & 305 \\
      \bottomrule
    \end{tabular}
    \begin{tablenotes}
      \small
      \item The columns are: (1) Observation date, (2) Modified Julian Date,
       (3) exposure time in seconds, (4) barycentric correction computed with {\tt
       rvsao}, and (5) the corrected radial velocity as measured with HeII 4686, with
       an error of 15 km/s as mainly from the uncertainties in the wavelength
       calibration.
    \end{tablenotes}
  \end{threeparttable}
\end{center}
\end{table}

\begin{table}
\begin{center}
  \begin{threeparttable}
    \caption{Properties of emission lines}
     \begin{tabular}{cllll}
      \toprule
      \toprule
        Line & FWHM & E.W. & Lum. & model \\
         ID  & (\AA) & (\AA) & $10^{34}$erg/s &  (\AA) \\
      \midrule
        HeII 4686 & 19.3 & 21.83 $\pm$ 0.20 & 43 & 21.75 \\
        HeI 5876 & 19.0 & 34.78 $\pm$ 0.29 & 49 & 34.21 \\
        HeI 6679 & 18.8 & 25.74 $\pm$ 0.37 & 24 & 26.56 \\
        HeII 5411 & 20.5 & 5.46 $\pm$ 0.13 & 8.3 & 6.10 \\
        HeI 4922 & 13.4 & 5.80 $\pm$ 0.64 & 8.4 & 3.91 \\
        HeI 4471 & 12.1 & 3.86 $\pm$ 0.65 & 7.0 & 5.18 \\
        $H_\gamma$ & 3.6 & 1.35 $\pm$ 0.22 & 2.7 \\
        $H_\beta$ & 4.5 & 7.51 $\pm$ 0.06 & 12 \\
        $H_\alpha$ & 4.7 & 26.54 $\pm$ 0.46 & 34 \\
        $[OIII]$ 4960 & 4.4 & 23.70 $\pm$ 0.49 & 40 \\
        $[NII]$ 6548 & 3.8 & 3.85 $\pm$ 0.39 & 4.7 \\
        $[NII]$ 6583 & 4.7 & 16.66 $\pm$ 0.08 & 18 \\
        $[SII]$ 6716 & 4.0 & 4.58 $\pm$ 0.07 & 4.0 \\
        $[SII]$ 6731 & 4.6 & 3.81 $\pm$ 0.06 & 3.1 \\
      \bottomrule

    \end{tabular}

    \begin{tablenotes}
      \small
      \item The columns are: (1) emission line ID, (2) FWHM as obtained from
        Gaussian fit, which equals to $2.35\sigma$, (3) equivalent width, (4) line
        luminosity in unit of $10^{34}$ erg/s, and (5) equivalent width from the best WR
        synthetic model.
    \end{tablenotes}
  \end{threeparttable}
\end{center}
\end{table}

%
%
%
%
%
%
%

\end{document}